\documentstyle{article}
\begin{document}
\title{Excitation of normal modes of a thin elastic plate
by moving dislocations}
\author{{\sc Rodrigo Arias and Fernando Lund} \\ 
Departamento de F\'\i sica, Facultad de Ciencias F\'\i sicas y  
Matem\'aticas \\ Universidad de Chile, Casilla 487-3, Santiago, Chile}
\date{}
\maketitle
\newcommand{\beq}{\begin{equation}}
\newcommand{\eeq}{\end{equation}}
\newcommand{\bea}{\begin{eqnarray}}
\newcommand{\eea}{\end{eqnarray}}

\begin{abstract}
We study the excitation
of harmonic waves in thin elastic samples by a single dislocation in 
arbitrary motion. We consider both screw and
edge dislocations that move perpendicularly to the surfaces of the layer. 
In Fourier space the displacement velocity  and dynamic stress 
fields generated by the motion of the dislocations are factored as
the product of two terms: one depends on the motion of the 
dislocation only, while the other is independent of it, and represents the 
medium's response. The latter term exhibits poles at frequencies that 
satisfy the dispersion relation of the harmonic modes of the plate. In the
case of a screw dislocation the modes that
are excited are a subfamily of the antisymmetric Rayleigh-Lamb  modes. 
For an edge dislocation a subfamily of the
symmetric Rayleigh-Lamb modes is excited, as well as the lowest lying
shear mode. The expression corresponding to a uniformly moving screw is 
worked out in detail; it has singular behavior at velocities coincident 
with the phase velocities of the allowed modes.
\end{abstract}

\section{Introduction}
Recent experimental results \cite{gro,sha1,sha2,sha3,bou,cil} 
in fast crack propagation
in brittle materials, such as glass and plexiglass, have provided evidence 
that crack propagation is dynamically unstable, and bounds 
that separate different crack behavior have been identified experimentally
in terms of the crack tip velocity. 
For example,
when the crack tip velocity surpasses $0.5 c_{R}$, with $c_{R}$ the 
Rayleigh wave velocity,  there are
strong sound emissions, oscillations of the crack tip velocity,  
and the surfaces left by the crack become quite rough. 

A scenario for a  
dynamic instability that could explain in a unified way  
the experimentally observed features of acoustic emissions, crack tip  
velocity oscillations and surface roughness has 
been recently proposed by one of us \cite{lun}. 
This scenario is based on continuum elasticity, and it incorporates  
the fact that a crack in motion can be modeled
as a continuous, time dependent, distribution of dislocations.  
These moving dislocations, when propagating in a medium that is itself  
oscillating, are subjet to two forces: one is the usual 
 Peach-Koehler force, and the second a
force that does no work \cite{lunr}, in this sense
analogous to the Lorentz magnetic force acting on a moving
charged particle. This latter force that does no work is proportional to the
dislocation's velocity, perpendicular to it,
and it is also proportional to the time derivative of the particle  
displacement generated by time dependent external loads.  
 The dynamic instability would occur by the following
mechanism: 1) the crack begins to propagate and starts emitting
elastic radiation; 2) normal modes of vibration of the plate are  
established, and they are responsible for a new oscillating force that does
no work and that is proportional to the crack tip velocity; 3) the
crack responds to this new periodic forcing by emitting radiation mainly
at the same frequencies at which it is being forced. This radiation 
reinforces the normal modes, that in turn act on the crack tip: this
is the instability.

The samples of brittle material that are used in 
experiments \cite{gro,sha1,sha2,sha3,bou,cil} are typically a few mm 
thick, and $\sim 20$ cm in width and length. Thus, it appears that useful 
insights into the underlying physics may be obtained by considering the 
samples as thin layers; that is, elastic layers of finite thickness and 
infinite in-plane dimensions. In this case the elastic sample is a wave 
guide whose possible horizontally polarized 
shear modes, and its mixed character
Rayleigh-Lamb modes are well known \cite{achenbach}.
In this work we have investigated the excitation of 
harmonic waves in such a wave guide by a dislocation, perpendicular to 
the surfaces of the layer, in arbitrary motion.
This represents a first step towards a firm analytic foundation  
for the 
scenario proposed in the previous paragraph.
The principal results of this work
are expressions, in Fourier space, that describe the displacement
velocity field and the stresses due to screw or edge dislocations moving
perpendicularly to the surfaces of the plate with arbitrary velocity,  
and the identification of the 
specific wave modes excited in each case. The relevant expressions  
are given as convolutions of a source function localized at the 
dislocation position with the medium's  
response, of which the Rayleigh-Lamb modes are poles.

In order to proceed, we use a formulation reported 
elsewhere \cite{rod} for the stress and velocity generated by a moving 
dislocation in finite elastic samples. This formulation is a 
generalization of line integral representations first found by Mura
\cite{mur} 
in the case of an infinite medium. The specialization of these results 
for a thin plate is given in Section 2. Section 3 contains the 
determination of amplitudes, whose precise definition is given in  
the text, associated with each possible harmonic mode. Expressions for 
particle velocity in terms of those amplitudes and a source function 
associated with the moving dislocation are given in Section 4 for a 
screw and in Section 5 for an edge dislocation. Section 6 discusses the 
corresponding stresses. Section 7 carries these results further in the 
special case of a uniformly moving dislocation, with the well known 
result of a dislocation in uniform motion in an infinite medium carried 
along for comparison. A discussion of the implications of these results 
for our current understanding of the brittle fracture of thin plates is 
offered in Section 8, and Section 9 has concluding remarks. The Fourier 
components of the Green's function of a thin plate 
\cite{chi} are written down 
explicitely in an Appendix.

\section{Displacement velocity due to a dislocation loop moving in a finite
sample} 
 
We consider small displacements $U_{m}(\vec{x},t)$ in  
a homogeneous elastic medium of density $\rho$ and elastic constants 
$C_{ijkl}$ \cite{achenbach}. A general formula for  
the velocity and stress fields generated by a dislocation loop  
undergoing arbitrary motion in an infinite medium 
was obtained some years ago by
Mura \cite{mur}. It is written 
as a convolution of  
the medium's impulse  
response with a source localized along the dislocation loop  
(i.e. independent of the slip plane).
These formulae have been generalized to the case of a bounded elastic
medium, as well as the formula for the displacement
field that involves integration over succesive slip planes
\cite{rod}. The
latter expression for the displacement field is:
\begin{equation}
U_{m}(x,t) = -b_{i} \int_{-\infty}^{\infty} dt' \int_{S(t')}
dS_{j}' \sigma_{ij}^{Gm}(x',x;t-t')
\: ,
\label{umf}
\end{equation} 
(attention should be paid to the order of the variables) where
$\sigma_{ij}^{Gm}(x,x';t-t')$ is the elastic stress associated
with the Green's function:
\beq
\sigma_{ij}^{Gm}(x,x';t-t') \equiv C_{ijkl} \frac{
\partial }{\partial x_{k}} G_{lm}(x,x';t-t') \: .
\eeq
Eq. (\ref{umf})
 is the analog of Mura's relation \cite{mur}. Since the problem is  
homogeneous under time translations it is possible to obtain an  
expression for particle velocity, the time derivative of particle  
displacement $U_m$ that involves an integral along the dislocation loop  
only. 
Indeed, since $\partial G_{mk}/\partial t = -\partial G_{mk}/\partial t'$,  
the displacement velocity becomes:
\beq
\frac{\partial U_{m}}{\partial t}(x,t)  =
b_{i} \int_{-\infty}^{\infty} dt' \{ \frac{\partial}{\partial t'} [
\int_{S(t')} dS_{j}'  \sigma_{ij}^{Gm}
(x',x;t-t') ]
 - \int_{\frac{dS}{dt'}(t')}
dS_{j}' \sigma_{ij}^{Gm}(x',x;t-t') \}
\: .
\eeq
The first term is zero since the Green's function vanishes at
$t-t'=\pm \infty$.
Also, since $\int_{dS(t')/dt'}dS_{j}' = \epsilon_{jpq} \int_{L(t')} dl_{q}'
V_{p}(x',t')$, where $L(t')$ is the dislocation loop bounding the slip
plane $S(t')$, $V_p(x',t')$ is the loop's local
 velocity and $\epsilon_{jpq}$ the completely
antisymmetric tensor in three dimensions, we have:
\begin{equation}
\frac{\partial U_{m}}{\partial t}(x,t) = -b_{i}
\int_{-\infty}^{\infty} dt'
\int_{L(t')} dl_{q}' \sigma_{ij}^{Gm}(x',x;t-t') \epsilon_{jpq}
V_{p}(x',t') \: .
\label{dudt}
\end{equation}
This expression for the particle velocity generated by a single 
dislocation loop within a finite elastic sample can  
also be obtained as a special case of the formulation of Chishko \cite{chi}  
for the particle velocity generated by a continuous distribution of  
dislocations in a thin elastic plate. 
 
We now specialize to the case of an infinite plate of thickness $2h$,  
with plane boundaries located at $z=\pm h$. 
The appropriate Green's function was calculated by Chishko \cite{chi} 
and it has the form:
\begin{equation}
G_{im}(z,z';\vec{R}-\vec{R}';t-t')=G_{im}^{\infty}(\vec{x}-\vec{x}';t-t')
-H_{im}(z,z';\vec{R}-\vec{R}';t-t') \: ,
\label{gim}
\end{equation}
with $G_{im}^{\infty}$ the Green's function 
appropriate for an infinite isotropic medium, 
and $H_{im}$ a kernel that cancels the normal 
stresses due to $G_{im}^{\infty}$ at the free surfaces. 
$H_{im}$ is the following convolution:
\begin{eqnarray}
H_{im}(z,z';\vec{R}-\vec{R}';t-t') & = & \Theta (t-t')
\sum_{a=1,2} \int_{0}^{(t-t')} d\tau \int dS_{a}''
\nonumber \\ & & \times G_{in}^{P}(z,\xi^{a};\vec{R}-\vec{R}'';
t-t'-\tau) \sigma_{nz}^{\infty m}(\xi^{a}-z';
\vec{R}''-\vec{R}';\tau) \: ,
\nonumber \\
\label{him}
\end{eqnarray}
with $dS_{a}'' = \pm dxdy$ for $a=1,2$
or $z''= \pm h$; $\Theta$ is the step function, and
$\xi^{a} = \pm h$. Here $\sigma_{nz}^{\infty m}$ is the elastic  
stress associated with the infinite medium Green's function,  
and $G_{in}^{P}(z,\xi^{a};\vec{R}-\vec{R}';t-t')$
is the Green's function of the homogeneous boundary value problem of the
plate: it corresponds to the displacement in direction $i$ produced  
by a point force applied at the surface in direction $n$  
at position ($\xi^{a}$, $\vec{R}'$), and time $t'$:
\begin{equation}
\rho \frac{\partial^{2}}{\partial t^{2}} G_{in}^{P}(z,\xi^{a};
\vec{R}-\vec{R}';t-t') - C_{ijlm} \frac{\partial^{2}}{\partial x^{j}
\partial x^{l}}G_{mn}^{P}(z,\xi^{a};\vec{R}-\vec{R}';t-t')=0
\: ,
\label{eginp}
\end{equation}
with $\xi^{a}=\pm h$, and subject to the following boundary conditions at
the surfaces:
\begin{eqnarray}
\sigma_{iz}^{Pn}(\xi^{b},\xi^{a};
\vec{R}-\vec{R}';t-t') & = & C_{izlm}\frac{\partial}{\partial 
x^{l}}G_{mn}^{P}(\xi^{b},\xi^{a};
\vec{R}-\vec{R}';t-t') \nonumber \\  
 & = & \pm \delta^{ab}\delta_{in}
\delta(\vec{R}-\vec{R}')\delta(t-t') \: ,
\label{bginp}
\end{eqnarray}
where $\pm$ corresponds to $a=1,2$ or to the surfaces $z'=
\xi^{a}=\pm h$ respectively.

For convenience, we have reproduced the explicit expressions
for the Fourier transforms of $G_{mn}^{P}(z,\xi;\vec{R}-\vec{R}';
t-t')$, $g_{mn}^{P \omega}(z,\xi|\vec{k})$,
from Ref. \cite{chi} in an Appendix:
\beq
g_{mn}^{P \omega}(z,\xi|\vec{k}) \equiv
\int_{-\infty}^{\infty} dt \int d\vec{R} e^{-i \omega t+ i
\vec{k} \cdot \vec{R} } G_{mn}^{P}(z,\xi;\vec{R};t)
\: .
\eeq
 
\section{Harmonic modes of a thin plate}
Harmonic modes in a thin elastic layer have properties, such as 
polarization and dispersion relations, that are well 
known \cite{achenbach}. In particular, the dispersion relation can be 
read off from the poles 
 of the Fourier transform of the Green's 
function $G_{ij}^{P}$, $g_{ij}^{P\omega}(z|\vec{k})$
(see Eqs. (\ref{gijp})-(\ref{eq:last}) in the Appendix). Poles of 
$g_{ij}^{P \omega}(z|\vec{k})$ occur when $\Delta_{a}(k^{2},\omega^{2})=0$,
$\Delta_{s}(k^{2},\omega^{2})=0$ and when $\sinh (q_{t}h) = 0$, 
$\cosh (q_{t}h) = 0$, with $q_{l,t} \equiv 
\sqrt{k^{2}-\omega^{2}/c_{l,t}^{2}}$, and $c_l$ (resp. $c_t$) are the speed 
of longitudinal (resp. transverse) waves in bulk material.

The poles $\Delta_{a}(k^{2},\omega^{2})=0$ of 
$g_{ij}^{P \omega}(z,\xi|\vec{k})$
give rise to antisymmetric modes, and we now determine their amplitudes, 
which will be needed in the next section, 
by taking the limit:
\begin{equation}
(v_{A}^{j})_{i} = \lim_{\Delta_{a} \rightarrow 0} \Delta_{a}
g_{ij}^{P \omega} (z,\xi|\vec{k}) \: .
\end{equation}
In this way three linearly independent modes, labelled by ``$j$''
 are obtained. They are vectors whose components are labelled by ``$i$''.
These antisymmetric modes, of mixed character,
are for $j=\beta=1,2=x,y$ and $j=3=z$ respectively: 
\begin{eqnarray}
\vec{v}^{\beta}_{A}(z|\vec{k}) & = & ik_{\beta} 
[ (k^{2}+q_{t}^{2})\sinh (q_{l}h) \cosh (q_{t}z)
-2q_{l}q_{t}
\sinh (q_{t}h) \cosh (q_{l}z)] 
\hat{z} \nonumber \\ 
&  & + \{ \frac{k_{\alpha}k_{\beta}}{k^{2}}q_{t}[
 (k^{2}+q_{t}^{2}) \sinh (q_{l}h)
\sinh (q_{t}z) -2k^{2} \sinh (q_{t}h)
\sinh (q_{l}z) ] \nonumber  \\
&  &  \qquad + (\delta_{\alpha \beta}-\frac{k_{\alpha}k_{\beta}}{k^{2}})
[ \frac{(k^{2}+q_{t}^{2})^{2}}{q_{t}}
\sinh (q_{l}h) \sinh (q_{t}z) \nonumber \\
 & & \qquad \qquad  - 4q_{l}k^{2} \tanh (q_{t}h) \cosh (q_{l}h)
\sinh (q_{t}z) ] \} \hat{e}_{\alpha} 
\nonumber \\
\vec{v}_{A}^{z}(z|\vec{k}) & = & q_{l}[(k^{2}+q_{t}^{2})
\cosh (q_{t}h) \cosh (q_{l}z) - 2k^{2} \cosh (q_{l}h) \cosh (q_{t}z) ] \hat{z}
\nonumber \\
&  & - ik_{\alpha}[(k^{2}+q_{t}^{2})\cosh (q_{t}h) \sinh (q_{l}z) -
2q_{l}q_{t} \cosh (q_{l}h) \sinh (q_{t}z) ] \hat{e}_{\alpha}
\: .
\nonumber \\
\end{eqnarray}
A mode that is not linearly independent of the latter can be obtained as:
$\vec{v}_{A}^{L}(z|\vec{k}) \equiv -k_{\beta} \vec{v}_{A}
^{\beta}(z|\vec{k})$. It will be seen below that this is 
the mode excited by a moving screw dislocation:
\begin{eqnarray}
\vec{v}_{A}^{L}(z|\vec{k}) & = & -ik^{2}[(k^{2}+q_{t}^{2})
\sinh (q_{l}h) \cosh (q_{t}z)-2q_{l}q_{t} \sinh (q_{t}h) \cosh (q_{l}z)]
\hat{z} \nonumber \\ 
& + & q_{t} k_{\alpha} \hat{e}_{\alpha} [ 2k^{2}
\sinh (q_{t}h) \sinh (q_{l}z)-(k^{2}+q_{t}^{2}) \sinh (q_{l}h) \sinh (q_{t}z)] 
\: .
\label{vd}
\end{eqnarray}

In an analogous way, the amplitude of the symmetric modes corresponding to 
$\Delta_{s}(k^{2},\omega^{2})=0$ are obtained as:
\begin{equation}
(v_{S}^{j})_{i} \sim \lim_{\Delta_{s} \rightarrow 0} \Delta_{s}
g_{ij}^{P \omega} (z,\xi|\vec{k}) \: .
\end{equation}
It will also be seen below that the
  mode excited by a moving edge dislocation corresponds to $j=3=z$:
\begin{eqnarray}
\vec{v}_{S}^{z}(z|\vec{k}) & = & 
q_{l}[(k^{2}+q_{t}^{2})\sinh (q_{t}h) \sinh (q_{l}z) 
- 2k^{2} \sinh (q_{l}h) \sinh (q_{t}z) ] \hat{z}
\nonumber \\
& - & i k_{\alpha} [ (k^{2}+q_{t}^{2}) \sinh (q_{t}h) \cosh (q_{l}z) -
2q_{l}q_{t} \sinh (q_{l}h) \cosh (q_{t}z) ] \hat{e}_{\alpha}
 \: .
\label{vdsz}
\end{eqnarray} 
The shear normal modes are of transverse character and 
polarized parallel to the surfaces of the plate. The
symmetric shear modes can be
obtained  by multiplying $g_{ij}^{P}$ by $\sinh (q_{t}h)$
and taking the limit $\sinh (q_{t}h) = 0$, or $q_{t}h = i n \pi $, 
$n=1,2, \ldots$. Thus:
\begin{equation}
\vec{v}_{S}^{\beta}(z|\vec{k}) = k^{2}(\delta_{\alpha \beta}
-\frac{k_{\alpha}k_{\beta}}{k^{2}})\cos (\frac{n \pi}{h}z) \hat{e}_{\alpha} \: .
\end{equation}
The antisymmetric shear modes correspond to multiplying $g_{ij}^{P}
(z,\xi|\vec{k})$
 by $\cosh (q_{t}h)$, and then
 taking the limit $\cosh (q_{t}h) = 0$, or $q_{t}h =
i(2n+1)\pi /2 $, $n=1,2,\ldots$:
\begin{equation}
\vec{v}_{A}^{\beta}(z|\vec{k}) = k^{2}(\delta_{\alpha \beta}
-\frac{k_{\alpha}k_{\beta}}{k^{2}})
\sin (\frac{(2n+1) \pi}{2h}z) \hat{e}_{\alpha} \: .
\end{equation}
Note that the modes with dispersion relation $\omega = c_{l,t}k$ that 
occur in an infinite isotropic medium are not normal modes of the plate.
The lowest shear mode of the plate has dispersion relation $\omega =
c_{t} k$, but it is only polarized in the plane of the plate. 
 
\section{Displacement velocity due to a screw dislocation moving 
perpendicularly to the 
surfaces of the plate}

We consider a straight 
screw dislocation that is parallel to the $z$ axis, or perpendicular 
to the surfaces of the plate, with a Burgers vector $\vec{b}=b \hat{z}$, 
in arbitrary motion. The dislocation as a curve is described 
by the location of its points as a function of time: 
$\vec{X}(\sigma,\tau) =(X_1(\tau),X_2(\tau),\sigma)$, 
with $\tau$ the time, and $\sigma$ parametrizing the curve. 
However, the dislocation loop or integration curve $L(t')$ of Eq. (\ref{dudt})
should also include segments on the surfaces of the plate and one at infinity
 (that it is easily seen not to contribute), in order to close the loop. 
The segments on the surfaces of the plate do not contribute 
to the displacement velocity due to the boundary condition satisfied by the 
Green's function on these surfaces:
the indices $p$ and $q$ in Eq. (\ref{dudt}) are different from $z$ on the 
segments on the surface, and so, because of the term $\epsilon_{jpq}$, 
$j$ must be equal to $z$, thus $\sigma_{iz}^{Gm}(x',x;t-t')$ 
is evaluated at the surfaces ($x'=x_{S}'$) where
it vanishes.
The expression in Eq. (\ref{dudt}) for the displacement velocity is then:
\begin{equation}
V_{m}(\vec{x},t) = - b \int_{-\infty}^{\infty} d\tau \int_{-h}^{h} d
\sigma C_{z \alpha kl} \epsilon_{\alpha \beta} \frac{\partial X_{\beta}}
{\partial \tau} \frac{\partial}{\partial x_{l}'} 
G_{mk}(z,z';\vec{R}-\vec{R}';t-\tau) \: ,
\end{equation}
with the convention that greek indices take values $(1,2)$ or $(x,y)$, 
and $\epsilon_{12}=-\epsilon_{21}=1$,
$\epsilon_{11}=\epsilon_{22}=0$. Thus:
\begin{eqnarray}
V_{m}(\vec{x},t) & = & -b \rho c_{t}^{2} \epsilon_{\alpha \beta} 
\int_{-\infty}^{\infty} d\tau \frac{\partial X_{\beta}}{\partial \tau} 
\int_{-h}^{h} d\sigma \{ -\frac{\partial}{\partial x^{\alpha}}G_{mz}
(\vec{x},\vec{X}(\sigma,\tau);t-\tau) \nonumber \\ 
 & & + \frac{\partial}{\partial \sigma}
G_{m \alpha}(\vec{x},\vec{X}(\sigma,\tau);t-\tau) \} \: .
\label{eq:velgr}
\end{eqnarray}
The in-plane Fourier components of the displacement velocity, are 
defined as:
\begin{equation}
V_{m}^{\omega}(z|\vec{k}) = \int_{-\infty}^{\infty} dt \int d\vec{R}
e^{-i \omega t + i \vec{k} \cdot \vec{R} } V_{m}(\vec{R},z,t) \: .
\label{vfs}
\end{equation}
Replacing this into (\ref{eq:velgr}) leads to
\begin{equation}
V_{m}^{\omega}(z|\vec{k}) = - \epsilon_{\alpha \beta} 
I_{\beta}^{\omega}(\vec{k}) F_{m \alpha}^{\omega}(z|\vec{k}) \: ,
\label{vfrs}
\end{equation}
with
\begin{equation}
I_{\beta}^{\omega}(\vec{k}) \equiv \int_{-\infty}^{\infty}
d\tau e^{-i \omega \tau + i \vec{k} \cdot \vec{X}(\tau)}
\frac{\partial X_{\beta}}{\partial \tau}(\tau) \: ,
\label{ibe}
\end{equation}
a factor that depends on the velocity of the dislocation, and
$F_{m \alpha}^{\omega}(z|\vec{k})$ 
a factor that represents the medium's 
response:
\begin{equation}
F_{m \alpha}^{\omega}(z|\vec{k}) \equiv b \rho c_{t}^{2} \{
i k_{\alpha} \int_{-h}^{h} d\sigma g_{mz}^{\omega}(z,\sigma|\vec{k})
+\int_{-h}^{h} d\sigma \frac{\partial}{\partial \sigma} 
g_{m \alpha}^{\omega}(z,\sigma|\vec{k}) \} \: .
\end{equation}
After some algebra, the following expressions for 
$F_{m \alpha}^{\omega}(z|\vec{k})$ can be obtained:
\begin{eqnarray}
F_{z \alpha}^{\omega}(z|\vec{k}) & = & \frac{ik_{\alpha}b}{q_{t}^{2}}
\{ 1-i\frac{ (v_{A}^{L})^{\omega}_{z}
(z|\vec{k})}{\Delta_{a}(k^{2},\omega^{2})} \} 
\label{fza} \\
F_{\delta \alpha}^{\omega}(z|\vec{k}) & = &
\frac{k_{\alpha}b}{q_{t}^{2} \Delta_{a}(k^{2},\omega^{2})}
(v_{A}^{L})^{\omega}_{\delta}(z|\vec{k}) \: .
\label{fda}
\end{eqnarray}

These factors $F_{m \alpha}^{\omega}(z|\vec{k})$ 
have poles when $\Delta_{a}(k^{2},\omega^{2})=0$, 
which gives
the dispersion relation of the antisymmetric Rayleigh-Lamb modes of the plate. 
The terms associated with these poles are proportional to 
the Rayleigh-Lamb mode $\vec{v}_{A}^{L\omega}(z|\vec{k})$ introduced 
in Eq. (\ref{vd}).
This shows that the principal contribution 
(it comes from $\Delta_{a} \rightarrow 0$, or $\omega \sim \omega_{n}^{a}(
\vec{k})$: the dispersion relation of the antisymmetric
Rayleigh-Lamb modes) to the displacement
velocity field due to a screw dislocation moving perpendicularly to the 
surfaces of the plate comes from the excitation of the antisymmetric
Rayleigh-Lamb modes associated with $(\vec{v}_{A}^{L})
^{\omega}(z|\vec{k})$. When 
Fourier transforming back to configuration space it becomes clear 
that only the poles contribute, as they should. An explicitely worked out 
example is given in Section 7.

\section{Displacement velocity due to an edge dislocation moving 
perpendicularly to the surfaces of the plate}

We consider a straight edge dislocation parallel to the $z$ axis, 
or perpendicular to the surfaces of the plate, with a Burgers vector 
$\vec{b}=b_{x}\hat{x} +
b_{y} \hat{y} =b_{\gamma} \hat{e}_{\gamma}$, and moving arbitrarily. 
According to Eq.
(\ref{dudt}), for this edge dislocation the displacement velocity becomes:
\begin{equation}
V_{m}(\vec{x},t) = \left. 
- b_{\gamma}\epsilon_{\alpha \beta} C_{\gamma \alpha kl}
\int_{-\infty}^{\infty} d\tau \int_{-h}^{h} dz' \frac{\partial X_{\beta}}
{\partial \tau}  \frac{\partial}{\partial x_{l}'} G_{mk}(z,z';
\vec{R}-\vec{R}';t-\tau) \right|_{\vec{R}'=\vec{X}(\tau)} \: .
\end{equation}
Here again, there is no contribution from the segments of the
loop $L(t')$ that are on the surfaces of the plate and at infinity, for 
the same reasons that were given 
for a screw dislocation. Calculating the Fourier components 
of the displacement velocity, one obtains them as the product of two factors:
\begin{equation}
V_{m}^{\omega}(z|\vec{k}) = - \epsilon_{\alpha \beta} 
I_{\beta}^{\omega}(\vec{k}) P_{m \alpha}^{\omega}(z|\vec{k}) \: ,
\end{equation}
where $I_{\beta}^{\omega}(\vec{k})$ is the factor dependent on the 
dislocation's velocity, Eq. (\ref{ibe}), and $P_{m \alpha}^{\omega}
(z|\vec{k})$ is a 
factor that represents the medium response:
\begin{eqnarray}
P_{z \alpha}^{\omega}(z|\vec{k}) & = & \frac{A_{\alpha}(\vec{b},\vec{k},
\omega^{2})}{\Delta_{s}(k^{2},\omega^{2})}
(v_{S}^{z})^{\omega}_{z}
(z|\vec{k}) \\
P_{\gamma \alpha}^{\omega}(z|\vec{k}) & = & B_{\gamma \alpha}(\vec{b},
\vec{k},\omega^{2})+ \frac{ C_{\gamma \alpha}(\vec{b},
\vec{k},\omega^{2})}{q_{t}^{2}}+
 \frac{A_{\alpha}(\vec{b},\vec{k},
\omega^{2})}{\Delta_{s}(k^{2},\omega^{2})}
(v_{S}^{z})^{\omega}_{\gamma} 
(z|\vec{k}) \: ,
\end{eqnarray}
with:
\begin{eqnarray}
A_{\alpha}(\vec{b},\vec{k},
\omega^{2}) & \equiv & - \frac{(\gamma^{2}-2)}{q_{l}^{2}\gamma^{2}}
[(\gamma^{2}-2)k^{2}b_{\alpha}+2k_{\alpha}k_{\rho}b_{\rho}] \: , \\
B_{\gamma \alpha}(\vec{b},\vec{k},
\omega^{2}) & \equiv & i \frac{k_{\gamma}}{q_{l}^{2}} \{
\frac{(\gamma^{2}-2)}{\gamma^{2}}b_{\alpha}
+2 \frac{c_{t}^{2}}{\omega^{2}}k_{\rho}b_{\rho}k_{\alpha}
\} \: , \\
C_{\gamma \alpha}(\vec{b},\vec{k},
\omega^{2}) & \equiv & i (k_{\alpha}b_{\gamma}+\delta_{\alpha \gamma}
k_{\rho}b_{\rho}-2\frac{c_{t}^{2}}{\omega^{2}}k_{\rho}b_{\rho}k_{\alpha}
k_{\gamma} ) \: ,
\end{eqnarray}
and $\gamma^{2} \equiv (c_{l}/c_{t})^{2}$.
Note that the terms of $P_{m \alpha}^{\omega}(z|\vec{k})$ which have poles at
$\Delta_{s}(k^{2},\omega^{2})=0$ are proportional to the 
symmetric normal mode of the plate,
$(\vec{v}_{S}^{z})^{\omega}(z|\vec{k})$, of Eq. (\ref{vdsz}). 
There is also a pole at $q_{t}^{2}=0$ with non zero residue associated.
This shows that the principal contribution 
(it comes from $\Delta_{s} \rightarrow 0$, or $\omega \sim \omega_{n}^{s}(
\vec{k})$: the dispersion relation of the symmetric
Rayleigh-Lamb modes) to the displacement velocity 
comes from the excitation of the Rayleigh-Lamb mode 
$(\vec{v}^{z}_{S})^{\omega} 
(z|\vec{k})$, and also from the lowest lying
shear mode. As in the case of the screw dislocation, going 
back to configuration 
space shows, via Cauchy's theorem, that only those modes contribute.

\section{Dynamic stresses on the plate}
Since the elastic stresses satisfy:
\begin{equation}
\frac{\partial}{\partial t} \sigma_{ij} = C_{ijlm} \frac{\partial V_{m}}
{\partial x_{l}} (\vec{x},t) \: ,
\end{equation}
for $\omega \neq 0$ its Fourier components are:
\begin{equation}
\sigma_{ij}^{\omega}(z|\vec{k}) = \frac{C_{ijlm}}{i \omega} \nabla_{l}
V_{m}^{\omega}(z|\vec{k}) \: ,
\label{sijw}
\end{equation}
with $\nabla_{l} \equiv (-ik_{x},-ik_{y},\partial/\partial z )$. We 
shall check that this expression satisfies the appropriate boundary 
conditions at the free surfaces.

\subsection{Screw dislocation}
Using Eqs. (\ref{vfrs}), (\ref{sijw}), 
one obtains the following expression for the stress produced
by a dislocation that moves perpendicularly to the surfaces of the plate
($\omega \neq 0$):
\begin{equation}
\sigma_{mz}^{\omega}(z|\vec{k}) = -\frac{1}{i \omega} 
I_{\beta}^{\omega}(\vec{k}) \epsilon_{\alpha \beta} \kappa_{m z}^{
\alpha \omega}
(z|\vec{k}) \: ,
\end{equation}
with:
\begin{eqnarray}
\kappa_{\alpha \omega}^{z z}(z|\vec{k}) & \equiv & 
\rho c_{t}^{2} \{ -i k_{\gamma}
(\gamma^{2}-2)F_{\gamma \alpha}^{\omega}(z|\vec{k}) +
\gamma^{2} \frac{\partial}{\partial z} F_{z \alpha}^{\omega}(z|\vec{k}) 
\} \nonumber \\
\kappa_{\delta z}^{\alpha \omega}(z|\vec{k}) & \equiv &  \rho c_{t}^{2}
\{ -i k_{\delta} F_{z \alpha}^{\omega}(z|\vec{k})+ \frac{\partial}{\partial z} 
F_{\delta \alpha}^{\omega}(z|\vec{k}) \} \: .
\end{eqnarray}
Using the expressions obtained for $F_{m \alpha}^{\omega}(z|\vec{k})$
(see Eqs. (\ref{fza}), (\ref{fda})), one
obtains:
\begin{equation}
\kappa_{\delta z}^{\alpha \omega}(z|\vec{k})  \equiv  b \rho c_{t}^{2}
\frac{k_{\delta}k_{\alpha}}{q_{t}^{2}} \{ 1 - \frac{1}{\Delta_{a}}
[(k^{2}+q_{t}^{2})^{2}\sinh (q_{l}h) \cosh (q_{t}z) -4 q_{l}q_{t}k^{2}
\sinh (q_{t}h)\cosh (q_{l}z) ] \} \: .
\label{kda}
\end{equation}
If one evaluates the stress $\sigma_{\gamma z}^{\omega}$ on the surfaces of the 
plate $z=\pm h$, one sees, using Eq. (\ref{kda}), that it vanishes, as it
should for a traction free surface. Also,
\begin{equation}
\kappa_{z z}^{\alpha \omega}(z|\vec{k}) = b \rho c_{t}^{2}
\frac{2ik_{\alpha}k^{2}(k^{2}+q_{t}^{2})}{q_{t}\Delta_{a}} \{
\sinh (q_{t}h) \sinh (q_{l}z)-\sinh (q_{l}h) \sinh (q_{t}z) \}
\: .
\end{equation}
This is also zero on the surfaces of the plate $z= \pm h$, as it should.
The other components of the stress also 
decompose as the product of two factors.

\subsection{Edge dislocation}
The Fourier components of the stress, $\sigma_{mz}^{\omega}(z|\vec{k})$,
produced by an edge dislocation moving perpendicularly to the surfaces of 
the plate ($\omega \neq 0$) are decomposed also into two factors:
\begin{equation}
\sigma_{mz}^{\omega}(z|\vec{k}) = -\frac{1}{i \omega}
\epsilon_{\alpha \beta}I_{\beta}^{\omega}(\vec{k}) Q_{m z}^{\alpha \omega}
(z|\vec{k}) \: ,
\end{equation}
with:
\begin{eqnarray}
Q_{z z}^{\alpha \omega}(z|\vec{k}) & = & D_{\alpha}(\vec{k},
\vec{b}) 
\{ 1-\frac{1}{\Delta_{s}
(k^{2},\omega^{2})}[(k^{2}+q_{t}^{2})^{2}\sinh (q_{t}h) \cosh (q_{l}z)
\nonumber \\ & & 
-4k^{2}q_{l}q_{t}\sinh (q_{l}h) \cosh (q_{t}z) ] \} \: , \\
Q_{\gamma z}^{\alpha \omega}(z|\vec{k}) & = & 
\frac{2ik_{\gamma}(k^{2}+q_{t}^{2})}{q_{l}\Delta_{s}(k^{2},
\omega^{2})} D_{\alpha}(\vec{k},\vec{b})
\{ \sinh (q_{t}h) \sinh (q_{l}z)-\sinh (q_{l}h) \sinh (q_{t}z) \} \: ,
\nonumber \\
\end{eqnarray}
and:
\begin{equation}
 D_{\alpha}(\vec{k},
\vec{b}) \equiv
\rho c_{t}^{2} \frac{(\gamma^{2}-2)}{\gamma^{2}}
[(\gamma^{2}-2)k^{2}b_{\alpha}+2k_{\alpha}k_{\rho}b_{\rho}] \: .
\end{equation}
These expressions for $Q_{\alpha \omega}^{m z}(z|\vec{k})$ assure that the
components of the stress $\sigma_{iz}^{\omega}$ are null on the surfaces of the 
plate.

\section{Example: a uniformly moving screw dislocation}
We consider the case of a screw dislocation that moves 
perpendicularly to the surfaces of a plate, with uniform 
velocity $V_{o} \hat{x}$. We will make a parallel
with the case of the same dislocation moving in an infinite medium, for which 
the particle velocity is well known \cite{hl}.
In the case of a thin layer, according to Eqs. (\ref{vfs}), (\ref{vfrs}), 
the displacement velocity in coordinate space is:
\begin{equation}
V_{m}(\vec{x},t) = -\frac{1}{(2 \pi)^{3}} \epsilon_{\alpha \beta}
\int d\vec{k} \int d\omega e^{i \omega t -i \vec{k} \cdot \vec{x}}
I_{\beta}^{\omega}(\vec{k})F_{m \alpha}^{\omega}(z|\vec{k})\: .
\label{vmi}
\end{equation}
For a screw dislocation moving in a plate the factor $F_{m \alpha}^{\omega}
(z|\vec{k})$ was given in Eqs. (\ref{fza}), (\ref{fda}). 
In an infinite medium the analogous quantity is:
\begin{equation}
F_{m \alpha}^{\omega}(z|\vec{k})  =  \frac{ik_{\alpha}b}{q_{t}^{2}} 
\delta_{mz} \: ,
\end{equation}
which shows that, in this case, 
only shear modes polarized in the $z$ direction are excited, since
there is a pole at $q_{t}^{2}=0$ or $\omega=\pm c_{t}k$. Notice that in the
case of the plate there is an ``apparent'' pole at $q_{t}^{2}=0$ also, but its
residue is zero, meaning that ``infinite medium'' shear modes are not excited.

Using the definition of $I_{\beta}^{\omega}(\vec{k})$  
given in Eq. (\ref{ibe}),  
Eq. (\ref{vmi}) for  the displacement velocity field component $m=3=z$ in
the plate, call it $V_{z}^{P}(\vec{x},t)$, can be written as 
(we will concentrate on the component $m=3=z$, which is the only nonvanishing 
component in the infinite medium case):
\begin{equation}
V_{z}^{P}(\vec{x},t) = -\frac{bV_{o}\delta_{\beta x}}
{(2\pi)^{3}}\epsilon_{\alpha \beta}
\int d\vec{k} \int d\omega \int d\tau 
e^{i\omega (t-\tau)-i \vec{k} \cdot (\vec{x}-V_{o}\tau \hat{x})}
\frac{i k_{\alpha}}{q_{t}^{2}} 
\{  
1-\frac{i(v_{A}
^{L})^{ \omega}_{z}(z|\vec{k})}{\Delta_{a}(k^{2},\omega^{2})}
 \} \: .
\end{equation}
The integrand in 
this expression has poles at $\Delta_{a}(k^{2},\omega^{2})=0$. These 
poles give the dispersion relation of the antisymmetric Rayleigh-Lamb
modes of the plate, $\omega = \omega_{n}^{a}(k)+i \epsilon$. A small positive
imaginary part has been added in order to satisfy causality: if 
$\tau > t$ there are no contributions because one closes the contour of
the $\omega$ integral in the lower half plane, where there are no poles.
 When $t > \tau$ we do the integral
over $\omega$ by closing the contour on the upper half plane, and one
obtains the contribution from all the poles already mentioned:
\begin{equation}
V_{z}^{P}(\vec{x},t) = \left. \frac{bV_{o}}
{(2\pi)^{2}}\epsilon_{\alpha x}
\frac{\partial}{\partial x_{\alpha}}
\sum_{n} \int d\vec{k} \int_{-\infty}^{t} d\tau 
e^{i \omega_{n}(k)(t-\tau)-i \vec{k} \cdot (\vec{x}-V_{o}\tau \hat{x})}
\frac{(v_{A}^{L})^{\omega}_{z} (z|\vec{k}) }
{q_{t}^{2}\frac{\partial \Delta_{a}}{\partial \omega}  
(k^{2},\omega^{2})} \right|_{\omega =
\omega_{n}(k)}
 \: ,
\end{equation} 
while the analogous expression in the infinite medium case 
for the $z$ component of the displacement velocity field, 
$V_{z}^{\infty}(\vec{x},t)$, comes from
calculating the residues at the poles $\omega_{n}=\pm c_{t}k$. This
gives:
\begin{equation}
V_{z}^{\infty}(\vec{x},t) = -\frac{ibV_{o}c_{t}^{2}\epsilon_{\alpha x} }
{(2\pi)^{2}} \frac{\partial}{\partial x_{\alpha}}
\int d\vec{k} \int_{-\infty}^{t} d\tau e^{-i \vec{k} \cdot (\vec{x}-
V_{o}\tau \hat{x})} \{
\frac{e^{ikc_{t}(t-\tau)}}{2kc_{t}}-\frac{e^{-ikc_{t}(t-\tau)}}{2kc_{t}} \}
\: .
\end{equation}
The last two equations show that the solution for $V_{z}(\vec{x},t)$ can
be written as a superposition of contributions from different modes.
The integration over $\tau$ can be done, and it renders for the case of the
plate:
\begin{equation}
V_{z}^{P}(\vec{x},t) = -\frac{ibV_{o}}{(2\pi)^{2}}
\frac{\partial}{\partial y}
\sum_{n} \int d\vec{k} \left.
\frac{e^{-i \vec{k} \cdot (\vec{x}-V_{o}t \hat{x})} }{(\omega_{n}(k)
-k_{x}V_{o})}
\frac{(v_{A}
^{L})_{z}^{\omega}
(z|\vec{k})}{q_{t}^{2}\frac{\partial}{\partial \omega} 
\Delta_{a}(k^{2},\omega^{2})} \right|_{\omega = \omega_{n}(k)} \: ,
\end{equation}
and in the case of the infinite medium:
\begin{equation}
V_{z}^{\infty}(\vec{x},t) = -\frac{bV_{o}
c_{t}}{2(2\pi)^{2}} \frac{\partial}{\partial y}
\int \frac{d\vec{k}}{k}  e^{-i \vec{k} \cdot (\vec{x}-
V_{o}t \hat{x})} \{
\frac{1}{(kc_{t}-k_{x}V_{o})}-\frac{1}{(-kc_{t}-k_{x}V_{o})} \}
\: . \label{vzin}
\end{equation}
These last two equations show that 
there would be a large contribution to the final result 
from the region where the ``Doppler shifted'' frequency of the  
modes $(\omega_{n}(k)-k_{x}V_{o})$ is
approximately zero (if that is possible for a given $V_{o}$). 
For example, in the infinite medium case if the velocity $V_{o}$ 
of the dislocation goes beyond $c_{t}$, one gets Cerenkov type 
radiation. But, for the case of the plate, the phase velocities
$\omega_{n}(k)/k_{x}$ are always greater than $V_{R}$, the 
Rayleigh wave velocity, meaning that there is no
large excitation of modes at uniform velocities $V_{o} < V_{R}$.

The angular integral over $\vec{k}$ can't be done analytically for the
case of the plate. We continue with the infinite medium case 
in order to illustrate
that one can get the exact solution as a superposition over the shear modes
excited. If $\vec{k}=k(\cos \theta,\sin \theta)$, then Eq. (\ref{vzin}) can
be written:
\begin{equation}
V_{z}(\vec{x},t) = \frac{ib
c_{t}}{2(2\pi)^{2}}
\int_{0}^{\infty} dk \int_{-\pi}^{\pi} d\theta \sin \theta
  e^{-i k (x-V_{o}t) \cos \theta -i k y \sin \theta } \{
\frac{1}{(\frac{c_{t}}{V_{o}}-\cos \theta)}+\frac{1}{(\frac{c_{t}}{V_{o}}+
\cos \theta)} \}
\: . 
\end{equation}
Doing the integral over $k$:
\begin{equation}
V_{z}(\vec{x},t) = \frac{b
c_{t}}{2(2\pi)^{2}} 
 \int_{-\pi}^{\pi} d\theta \frac{\sin \theta}
{((x-V_{o}t) \cos \theta + y \sin \theta )} \{
\frac{1}{(\frac{c_{t}}{V_{o}}-\cos \theta)}+\frac{1}{(\frac{c_{t}}{V_{o}}+
\cos \theta)} \}
\: . 
\end{equation}
This angular integral can be done by using the change of
variables $z=e^{i \theta}$, then it becomes an integral over the
unit circle in the complex plane, and then one applies Cauchy's theorem. 
The result is:
\begin{equation}
V_{z}(\vec{x},t) = \frac{bV_{o}}{2 \pi} \frac{y \sqrt{1-(\frac{V_{o}}
{c_{t}})^{2}}}{[(x-V_{o}t)^{2}+y^{2}(1-(\frac{V_{o}}{c_{t}})^{2}) ] }
\: ,
\end{equation}
which corresponds to the exact solution that can be obtained in a simpler
way \cite{hl}:
\begin{equation}
U_{z}(\vec{x},t) = \frac{b}{2\pi} \tan^{-1} ( \frac{y \sqrt{1-(\frac{V_{o}}
{c_{t}})^{2}} }{(x-V_{o}t)} ) \: .
\end{equation}
This is, of course, 
the static solution for a screw dislocation in an infinite
medium, $U_{z}=(b/2\pi)\phi=(b/2\pi)\tan^{-1}(y/x)$,
when Lorentz-transformed to a frame moving with velocity $V_0$.

\section{Relation with experiments on fast fracture}
Experiments on fast fracture 
of glass and plexiglass \cite{gro,sha1,sha2,sha3,bou,cil} 
have been made in 
mode I loading, which means that, in order to use our results, 
the crack should be modeled by edge dislocations.
Our results about the motion of edge dislocations show that the 
mode excited in this case is the 
Rayleigh-Lamb mode associated with $(\vec{v}_{S}^{z})^{\omega}
(z|\vec{k})$. 
Moreover, experiments exhibit acoustic and surface roughness spectra 
that have strong peaks at frequencies corresponding to the lowest lying 
long wavelength Rayleigh-Lamb modes \cite{lun},
and it is tempting to think that long wavelength modes, requiring less 
energy, will be more efficiently excited.
The long wavelength solution 
for the dispersion relation of these modes comes from taking the limit 
$k \rightarrow 0$ in $\Delta_{s}(k^{2},\omega^{2})=0$, 
and leads to two possibilities: $\cos (\omega h/c_{l})=0$ 
or ``odd'' frequencies $\omega_{n} = (2n+1)\pi c_{l}/2h$ ($n=1,2, \ldots$) and 
$\sin (\omega h/c_{t})=0$ or ``even'' frequencies
$\omega_{m} = m \pi c_{t}/h$ ($m=1,2,\ldots$).
The limit $k \rightarrow 0$
for the mode $\vec{v}_{S}^{z}(z|\vec{k})$ is:
\begin{equation}
\vec{v}_{S}^{z}(z|\vec{k}) \simeq -i \frac{\omega^{3}}
{c_{l}c_{t}^{2}} \sin (\frac{\omega}{c_{t}}h) \sin (\frac{\omega}{c_{l}}z)
\hat{z}
\end{equation}
This expression goes to zero for the frequencies previously called ``even'', 
which means that the amplitude of excitation of these modes
will be lower than the corresponding amplitudes associated with the
"odd" frequencies. The fact that experiments deal with a crack, that is 
with many dislocations side by side, rather than a single dislocation 
will not change this fact and it may well happen that deviations from 
straight-front crack tip behaviour must be considered.

\section{Concluding remarks}
In this paper we have given an expression for the elastodynamic fields 
that are excited in a thin plate by a dislocation, perpendicular to the 
faces of the plate, in arbitray motion. This expression 
is a convolution of the 
impulse response of the plate with a source function that is localized 
at the dislocation position. An explicit expression for the impulse 
response has been determined in Fourier space as a rational function, 
with poles at the well known Rayleigh Lamb modes, both for screw and 
edge dislocations. Application to the case of a uniformly moving screw 
shows that large amounts of radiation are to be expected when the 
dislocation's velocity coincides with the phase velocity of one of the 
modes.

This work is part of an ongoing project that attempts to understand 
recently observed dynamic crack instabilities in thin plates 
in terms of the interaction 
of the crack tip with the wave modes of the plate. We hope the formalism 
presented in this work will prove helpful in advancing this project.

\section{Acknowledgments}
We thank J. F. Boudet and S. Ciliberto for communicating their experimental 
results previous to publication. This work was supported in part by 
Fondecyt Grants 3950011 and 1960892, Fundaci\'on Andes, 
and by a C\'atedra Presidencial en 
Ciencias.

\section*{Appendix: Fourier components of the plate's Green's function}
In this Appendix we reproduce the expression for the Green's function of 
a thin plate given by Chishko \cite{chi}.
According to Eq. (\ref{gim}) 
the Fourier components of the plate's Green's function are:
\begin{equation}
g_{im}^{\omega}(z,z'|\vec{k})=g_{im}^{\infty \omega}(z-z'|\vec{k})-
h_{im}^{\omega}(z,z'|\vec{k}) \: .
\end{equation}
From Eq. (\ref{him}):
\begin{eqnarray}
h_{im}^{\omega}(z,z'|\vec{k}) & =  & \sum_{a} (\pm) [-i k_{\beta}C_{nz \beta p}
g_{in}^{P \omega}(z,\xi^{a}|\vec{k})g_{pm}^{\infty \omega}(\xi^{a}
-z'|\vec{k}) 
\nonumber \\ & + & C_{nzzp}g_{in}^{P \omega}(z,\xi^{a}|\vec{k})
\frac{\partial}{\partial z} g_{pm}^{\infty \omega}(\xi^{a}-z'|\vec{k}) ]
\: ,
\end{eqnarray}
where ($\pm$) refers to $a=1,2$ or $\xi^{a}=\pm h$. The Fourier transforms
 of 
the
infinite medium's Green's function are \cite{chi}:
\begin{equation}
g_{ij}^{\infty \omega }(z-z'|\vec{k}) = -\frac{1}{2 \rho \omega^{2}} \{
s_{i}^{l}s_{j}^{l}\frac{e^{-q_{l}|z-z'|}}{q_{l}} -
[s_{i}^{t}s_{j}^{t}+\frac{\omega^{2}}{c_{t}^{2}} \delta_{ij}]
\frac{e^{-q_{t}|z-z'|}}{q_{t}} \} \: ,
\end{equation}
with $\vec{s}^{l,t} \equiv (ik_{x},ik_{y},q_{l,t}s(z-z'))$, 
$s(x) \equiv sign(x)$, and $q_{l,t} \equiv 
\sqrt{k^{2}-\omega^{2}/c_{l,t}^{2}}$. The Fourier transforms of the Green's
function $G_{ij}^{P}$ (see Eqs. (\ref{eginp}), (\ref{bginp})) are \cite{chi}:
\begin{equation}
g_{ij}^{P \omega}(z,\xi|\vec{k}) = \frac{1}{2 \rho c_{t}^{2}} 
\sum_{\lambda =l,t} M_{ij}^{\lambda \omega}(z,\xi|\vec{k}) \: .
\label{gijp}
\end{equation}
Here $M_{ij}^{\lambda \omega}(z,\xi|\vec{k})$ are matrices with components:
\begin{eqnarray}
M_{zz}^{l \omega}  =  q_{l}(k^{2}+q_{t}^{2}) \Phi_{2}(z|\xi) & ; &
M_{zz}^{t \omega}  =  -2 q_{l} k^{2} \Phi_{2}(\xi|z) \nonumber \\
M_{z \beta}^{l \omega}  =  -2i q_{l}q_{t} k_{\beta} \Phi_{4}(z|\xi)
& ; & 
M_{z \beta}^{t \omega} = i(k^{2}+q_{t}^{2})k_{\beta} \Phi_{3}(\xi|z)
\nonumber \\
M_{\alpha z}^{l  \omega}  =  -i(k^{2}+q_{t}^{2})k_{\alpha} \Phi_{3}
(z|\xi) & ; &
M_{\alpha z}^{t \omega}  =  2i q_{l}q_{t} k_{\alpha} \Phi_{4}(\xi|z)
\nonumber \\
M_{\alpha \beta}^{l \omega}  =  -2 q_{t} k_{\alpha}k_{\beta} \Phi_{1}(z|\xi) 
& ; & \nonumber 
\end{eqnarray}
\beq  
M_{\alpha \beta}^{t  \omega}  =  \frac{k_{\alpha}k_{\beta}}
{k^{2}}q_{t}(k^{2}+q_{t}^{2}) \Phi_{1}(\xi|z)
+(\delta_{\alpha \beta}-\frac{k_{\alpha}k_{\beta}}{k^{2}})
[ \frac{(k^{2}+q_{t}^{2})^{2}}{q_{t}}\Phi_{1}(\xi|z) -4q_{l}k^{2}
\Psi (\xi|z) ]  \: 
\eeq
where $\xi$ is equal to either $h$ or $-h$. The functions 
$\Phi_{p}(u|v)$ and $\Psi(u|v)$ are defined as:
\begin{eqnarray}
\Phi_{1}(u|v) & = & \frac{1}{\Delta_{s}} \cosh (q_{l}u) \cosh (q_{t}v)
+\frac{1}{\Delta_{a}} \sinh (q_{l}u) \sinh (q_{t}v) \nonumber \\
\Phi_{2}(u|v) & = & \frac{1}{\Delta_{s}} \sinh(q_{l}u) \sinh (q_{t}v)
+\frac{1}{\Delta_{a}} \cosh (q_{l}u) \cosh (q_{t}v) \nonumber \\
\Phi_{3}(u|v) & = & \frac{1}{\Delta_{s}} \cosh (q_{l}u) \sinh (q_{t}v)
+\frac{1}{\Delta_{a}} \sinh (q_{l}u) \cosh (q_{t}v) \nonumber \\
\Phi_{4}(u|v) & = & \frac{1}{\Delta_{s}} \sinh (q_{l}u) \cosh (q_{t}v)
+\frac{1}{\Delta_{a}} \cosh (q_{l}u) \sinh (q_{t}v) \nonumber \\
\Psi (u|v) & = & \frac{1}{\Delta_{s}} \coth (q_{t}u) \sinh (q_{l}u)
\cosh (q_{t}v) \nonumber \\  
& & + \frac{1}{\Delta_{a}} \tanh (q_{t}u) \cosh (q_{l}u)
\sinh (q_{t}v) \: ,
\label{phis}
\end{eqnarray}
with:
\begin{eqnarray}
\Delta_{s}(k^{2},\omega^{2}) & = & 
(k^{2}+q_{t}^{2})^{2} \cosh (q_{l}h) \sinh (q_{t}h)
-4k^{2}q_{l}q_{t} \cosh (q_{t}h) \sinh (q_{l}h) \nonumber \\
\Delta_{a}(k^{2},\omega^{2}) & = & 
(k^{2}+q_{t}^{2})^{2} \sinh (q_{l}h) \cosh (q_{t}h)
-4k^{2}q_{l}q_{t} \sinh (q_{t}h) \cosh (q_{l}h)
\label{eq:last} 
\end{eqnarray}

\end{document}